\documentclass[aps,prl,reprint,groupedaddress]{revtex4-1}

\usepackage{mathptmx}

\usepackage{float}
\usepackage{mathrsfs}
\usepackage{graphicx,amsfonts,bm,amssymb,amsmath,braket,hyperref}
\hypersetup{colorlinks=true,allcolors=[rgb]{0,0,1}}

\begin{document}


\title{Bound-State Band Reconstruction and Resonance in Spin-1/2 Bose Gas with 1D Spin-Orbit Coupling}



\author{Qi Gu, Yuncheng Xiong, Lan Yin}
\email[]{yinlan@pku.edu.cn}


\affiliation{School of Physics, Peking University, Beijing 100871, China}


\date{\today}

\begin{abstract}
In this work, we study two-body bound states in two-component Bose gas with a one-dimensional (1D) spin-orbit coupling (SOC) induced by Raman lasers.  The finite Raman coupling strength generates coupling among three spin channels, resulting in the reconstruction of three bound-state bands. In addition, multiple resonances can be induced at finite scattering lengths.  By tuning the interaction in one intra-species channel, one bound-state band can be lifted and three resonances can be achieved at different center-of-mass momenta (COMM),  which can be observable under current experimental conditions in ${}^{87}$Rb atoms.
\end{abstract}



\maketitle

{\textit{Introduction}}.
Synthetic spin-orbit coupling (SOC) is an important tool in the study ultracold quantum gases. There have been a lot of experimental and theoretical studies on spin-orbit coupled quantum gases during the past decade~\cite{Review-Spielman,goldman2014light,zhai2015degenerate,review-qizhou}.
The 1D SOC was first generated by dressing two hyperfine spin states with a pair of lasers in a ${}^{87}$Rb Bose-Einstein condensate~\cite{Nature2011Spielman}. The similar scheme was also used to generate 1D SOC in ${}^{40}$K~\cite{PhysRevLett.109.095301} and ${}^{6}$Li~\cite{PhysRevLett.109.095302} Fermi gases. 2D SOC has been experimentally realized in ${}^{40}$K gases with three Raman lasers~\cite{2DSOC-F} and in ${}^{87}$Rb atoms with an optical Raman lattice~\cite{2DSOC@lattice}. Bose gases with 1D SOC can condense into stripe phase, magnetized phase, or non-magnetized phase for different SOC parameters~\cite{Nature2011Spielman,ji2014experimental,PhysRevLett.108.225301}. Fermi gases with SOC were predicted to be unconventional superfluid at low temperatures~\cite{PhysRevLett.107.195304,PhysRevLett.109.105303,PhysRevLett.110.110401}.
Condensation of two-body bound states were predicted not only in Fermi gases~\cite{PhysRevLett.107.195305,PhysRevB.84.014512}, but also in Bose gases with SOC~\cite{Li}.

How SOC affects bound-state formation and atom scattering generally in a Bose gas with SOC is an important and unanswered problem. Interactions between atoms can be strongly altered by the light dressed with~\cite{williams2012synthetic,PhysRevB.83.094515}. Previous studies showed that in the case of fermions with 1D SOC, finite Raman strength can shift the location of the Feshbach resonance~\cite{PhysRevB.83.094515,ZhangPeng,PhysRevLett.111.095301,PhysRevA.94.053635}.  In a Bose gas with anisotropic SOC, one can induce resonance by tuning the anisotropy of SOC strengths~\cite{Gu}.  In the case with Rashba SOC, the resonance position can only be shifted in intra-species channels~\cite{Li}.  In the case with Wely SOC, resonance positions of all three scattering channels are shifted~\cite{Luo}.  Nonetheless, in the case of 1D SOC with vanishing Raman coupling, the SOC does not change either the resonance position or bound-state binding energy.

In this work, we study two-body bound states in spin-1/2 Bose gas with 1D SOC at finite Raman coupling.  The Raman coupling can be viewed as the effective zeeman field that causes spin-flipping processes.   We find that three scattering channels are coupled together resulting in the formation of three new bound-state bands.  The finite Raman coupling also induces resonances at finite scattering lengths.  By tuning the scattering length in one intra-species channel, one bound state band can be lifted up and the resonance locations can be shifted.  We propose a scheme to observe this resonance in ${}^{87}$Rb system.

\textit{Model}.
We consider a two-component homogeneous Bose gas with a Raman-induced SOC, described by the Hamiltonian $H=H_0+H_\text{int}$.  The single-particle term is given by
\begin{equation}
H_0=\epsilon_k + \dfrac{\Omega}{2} \sigma_x+(\dfrac{\hbar^2k_0}{m}k_x+\dfrac{\delta}{2})\sigma_z ,
\end{equation}
where $\Omega$ is the Raman coupling strength, $k_0$ is the SOC strength, and $ \delta $ is the detuning energy, $ \sigma_x $ and $ \sigma_z $ are Pauli matrices, and $\epsilon_k=\hbar^2 k^2/(2m)$. The recoil energy is defined as $ E_r=\hbar^2k_0^2/(2m) $.  The single-particle Hamiltonian $H_0$ can be diagonalized, yielding two helical excitation branches $\varepsilon_{\mathbf{k}}^{\pm}=\epsilon_\mathbf{k}\pm \sqrt{\Omega^2/4+(\hbar^2k_0k_x/m+\delta/2)^2}$. For simplicity, we just consider the case with zero detuning. The energy minimum of the lower branch $\varepsilon_{\mathbf{k}}^-$ is given by
\begin{equation}\label{cases}
E_\text{min}=
\begin{cases}
\epsilon_{\pm \mathbf{q}_0}^{-}=-E_r-\Omega^2/(16E_r), & \Omega<4E_r\\
\epsilon_0^{-}=-\Omega/2, & \Omega>4E_r
\end{cases},
\end{equation}
where $ \mathbf{q}_0=\mathbf{k}_0\sqrt{1-\Omega^2/(4E_r)^2}$ and $\mathbf{k}_0=k_0\hat{x}$.
The spin-dependent $ s $-wave interactions between bosons are given by
\begin{equation}
	H_\text{int}=\dfrac{1}{2 V} \sum_{\mathbf{k k' q} \rho \rho'} g_{\rho \rho'} c_{\mathbf{q+k'} \rho}^{\dagger} c_{\mathbf{q-k'} \rho'}^{\dagger} c_{\mathbf{q-k} \rho'} c_{\mathbf{q+k} \rho},
\end{equation}
where $ c^\dagger_{\mathbf{q}\rho} $ is the creation operator of a boson with momentum $ \hbar\mathbf{q} $ and spin component $ \rho=\uparrow$ or $\downarrow$. The $s$-wave coupling constant $ g_{\rho\!\rho'}$ is related to the scattering length in the absence of SOC $a_{\rho\rho'}$ by the renormalization relation $ 1/g_{\rho\rho'}=1/g_{\rho'\rho}=m(4\pi\hbar^2 a_{\rho\rho'})^{-1} -\Lambda$ with $\Lambda=\int d^3k (2\pi)^{-3} (2\epsilon_k)^{-1}$.

\textit{Two-body bound state}.
The eigenequation of a two-body bound state is given by  $H\ket{\varphi} = E_{2\mathbf{q}} \ket{\varphi}$, where $E_{2\mathbf{q}}$ and $\ket{\varphi}=\frac{1}{2}\sum_{\mathbf{k}\rho\rho'}\phi_{\rho\rho'}(\mathbf{q},\mathbf{k})c_{\mathbf{q}+\mathbf{k}\rho}^\dagger c_{\mathbf{q}-\mathbf{k}\rho'}^{\dagger}\ket{0} $ are bound state eigenenergy and eigenstate with COMM $2\mathbf{q}$.
From the eigenequation, we obtain a set of linear equations for the coefficient $\phi_{\rho\rho'}(\mathbf{q},\mathbf{k})$

\begin{equation}\label{Eq4}
	M_\mathbf{kq}\Phi_\mathbf{kq}=\mathbf{G}\frac{1}{V}\sum_{\mathbf{p}}\Phi_\mathbf{pq},
\end{equation}
where
$
\Phi_\mathbf{kq}=\left[\phi_{\uparrow\!\uparrow}(\mathbf{q},\mathbf{k}), \phi_{\downarrow\!\downarrow}(\mathbf{q},\mathbf{k}), \phi_{\uparrow\!\downarrow}(\mathbf{q},\mathbf{k}), \phi_{\downarrow\!\uparrow}(\mathbf{q},\mathbf{k})\right]^\intercal ,
$
and the matrix $ M_\mathbf{kq} $ is given by
\begin{equation*}\label{M}
\begin{pmatrix}
\xi_{\mathbf{kq}}-\delta-2h_{qx} & 0 & -\Omega/2 & -\Omega/2\\
0 		 & \xi_{\mathbf{kq}}+\delta+2h_{qx} & -\Omega/2 & -\Omega/2\\
-\Omega/2 & -\Omega/2 & \xi_{\mathbf{kq}}-2h_{kx} & 0\\
-\Omega/2 & -\Omega/2 & 0 & \xi_{\mathbf{kq}}+2h_{kx}
\end{pmatrix}
\end{equation*}
with $ \xi_{\mathbf{kq}}=E_{2\mathbf{q}}-\epsilon_{\mathbf{k}+\mathbf{q}}-\epsilon_{\mathbf{k}-\mathbf{q}}$, $ h_{kx}=\hbar^2k_0k_x /m$, and the coupling matrix $\mathbf{G}$ is given by $ G_{ij}= (g_{\uparrow\!\uparrow}\delta_{i1} + g_{\downarrow\!\downarrow}\delta_{i2} + g_{\uparrow\!\downarrow}\delta_{i3} + g_{\uparrow\!\downarrow}\delta_{i4})\delta_{ij} $.
Eq. (\ref{Eq4}) can be further written as
\begin{equation} \Gamma=\mathbf{G}\frac{1}{V}\sum_{\mathbf{k}}M_\mathbf{kq}^{-1}\Gamma ,
\end{equation}
where $ \Gamma=\mathbf{G}\frac{1}{V}\sum_{\mathbf{p}}\Phi_\mathbf{pq} $. The eigenenergy $E_{2\mathbf{q}}$ can be determined from the secular equation
\begin{equation}\label{det}
\det\left(\mathbf{G}\dfrac{1}{V}\sum_{\mathbf{k}}M^{-1}_\mathbf{kq}-\mathbf{I}\right)=0 ,
\end{equation}

Eigenenergies of both two-boson and two-fermion bound states satisfy Eq.~\eqref{det}.
For two-boson bound state, Eq.~\eqref{det} can be further written as

\begin{equation}\label{eigenenergy}
\det\begin{pmatrix}
	A_1-{1}/{a_{\uparrow\!\uparrow}} & \sqrt{2}C_1 & C_3 \\
	\sqrt{2}C_1 & A_3-{1}/{a_{\uparrow\!\downarrow}} & \sqrt{2}C_2 \\
	C_3  & \sqrt{2}C_2 &  A_2-{1}/{a_{\downarrow\!\downarrow}}
\end{pmatrix}=0,
\end{equation}
where $ A_{1,2,3} $ and $ C_{1,2,3} $ are functions of energy $ E_{2\mathbf{q}} $, momentum $ 2\hbar\mathbf{q} $, and $ \Omega $, as given below,
\begin{align}\label{AC}
\nonumber \dfrac{mA_{1,2}}{4\pi\hbar^2}&=\frac{1}{V}\sum_{\mathbf{k}}\dfrac{(\xi_{\mathbf{kq}}\pm 2h_{qx})(\xi_{\mathbf{kq}}^2-4h_{kx}^2)-\xi_{\mathbf{kq}}\Omega^2/2}{(\xi_{\mathbf{kq}}^2-4h_{qx}^{2})(\xi_{\mathbf{kq}}^2-4h_{kx}^2)-\xi_{\mathbf{kq}}^2\Omega^2}+\Lambda,\\
\nonumber
\dfrac{mA_3}{4\pi\hbar^2}&=\frac{1}{V}\sum_{\mathbf{k}}\dfrac{\xi_{\mathbf{kq}}(\xi_{\mathbf{kq}}^2-4h_{qx}^{2})}{(\xi_{\mathbf{kq}}^2-4h_{qx}^{2})(\xi_{\mathbf{kq}}^2-4h_{kx}^2)-\xi_{\mathbf{kq}}^2\Omega^2}+\Lambda,\\
\nonumber
\dfrac{mC_{1,2}}{4\pi\hbar^2}&=\frac{1}{V}\sum_{\mathbf{k}}\dfrac{\xi_{\mathbf{kq}}(\xi_{\mathbf{kq}}\pm 2h_{qx})\Omega/2}{(\xi_{\mathbf{kq}}^2-4h_{qx}^{2})(\xi_{\mathbf{kq}}^2-4h_{kx}^2)-\xi_{\mathbf{kq}}^2\Omega^2},\\
\dfrac{mC_3}{4\pi\hbar^2}&=\frac{1}{V}\sum_{\mathbf{k}}\dfrac{\xi_{\mathbf{kq}}\Omega^2/2}{(\xi_{\mathbf{kq}}^2-4h_{qx}^{2})(\xi_{\mathbf{kq}}^2-4h_{kx}^2)-\xi_{\mathbf{kq}}^2\Omega^2}.
\end{align}
When the Raman strength $ \Omega $ is finite, all the off-diagonal matrix elements in Eq.~\eqref{eigenenergy} are finite, indicating that in the presence of Raman field, the three spin channels $ \uparrow\!\uparrow, \downarrow\!\downarrow $ and $ \uparrow\!\downarrow $ mix together.

\begin{figure}[]
	\includegraphics[width=\columnwidth]{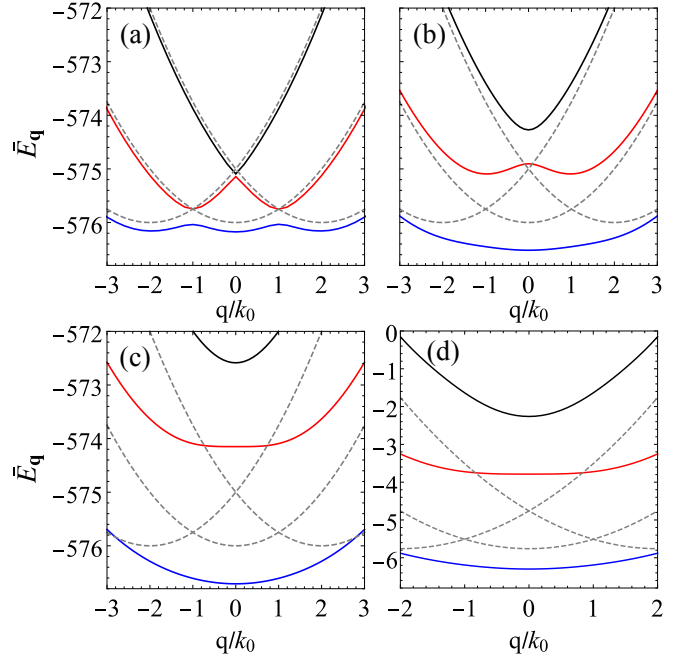}
\caption{\label{fig1} Energy bands of two-body bound states with symmetric interactions $a_{\uparrow\uparrow}=a_{\downarrow\downarrow}=a_{\uparrow\downarrow}=a$. The three parabolic dashed lines in each figure are energy bands at zero Raman strength as a reference, where the three spin channels are decoupled. The dimensionless energy $\bar{E}_\mathbf{q}$ is defined as $(E_\mathbf{q}-2E_{\min})/(2E_r) $. (a)  Energy bands at $\Omega=0.4E_r$ and $1/(k_0a)=24$, where the lowest band has three minima.  Formation of these new bands are due to couplings between spin channels. (b) Energy bands at $\Omega=2E_r$ and $1/(k_0a)=24$, where the lowest band has only one minimum.  (c) Energy bands at $\Omega=4E_r$ and $1/(k_0a)=24$, where two minima of middle band are merged.  (d) Energy bands at $\Omega=4E_r$ and $1/(k_0a)=2.4$, where all the bands are lifted up due to the increase in $ a $. }
\end{figure}

In the zero Raman strength limit, $ C_{1,2,3}=0$, Eq.~\eqref{eigenenergy} is reduced to three independent equations.  Three spin channels are decoupled, and the bound-state energy depends only on the scattering length of its spin channel, in the intra-species channel
\begin{equation}\label{X1X2}
\frac{E_{2\mathbf{q}}}{2E_r}=\frac{E_{\min}}{E_r}-\frac{1}{k_0^2a_{\rho\rho}^2}+(1\pm \frac{q_x}{2k_0})^2+(\frac{q_{yz}}{2k_0})^2,
\end{equation}
and in the inter-species channel
\begin{equation}\label{X3}
\frac{E_{2\mathbf{q}}}{2E_r}=\frac{E_{\min}}{E_r}-\frac{1}{k_0^2a_{\uparrow\!\downarrow}^2}+(\frac{q}{2k_0})^2.
\end{equation}
As shown in Fig.~\ref{fig1}, the minimum of the bound-state band in the intra-species $ \uparrow\!\uparrow $ ($\downarrow\!\downarrow$) channel is located at COMM  $ -2\hbar\mathbf{k}_0 $ ($ +2\hbar\mathbf{k}_0 $), while that in the inter-species $ \uparrow\!\downarrow $ channel is located at zero COMM. Bound states composed of two spin-1/2 atoms behave as a single spin-1 particle with a pure 1D SOC, $ \frac{\hbar^2k_0k_x}{m}F_z $, where $ F_z $ is the $z$-component spin operator for the spin-1 bound state.

When Raman coupling strength is finite, the three parabolic bands are reconstructed to three new disjoint energy bands.  As shown in Fig.~\ref{fig1}(a), with symmetric interactions $a_{\uparrow\!\uparrow}=a_{\downarrow\!\downarrow}=a_{\uparrow\!\downarrow}=a$, the lowest band has three minimum points located near $\pm 2\mathbf{k}_0$ and $ 0 $ when $\Omega$ is very small. In such case, to the first order of $ \Omega $, bound states can be approximated as a spin-1 single particle with SOC described in Ref.~\cite{Ohberg-spin-1}. As $\Omega$ increases, two minimum points in the lowest band disappear and only the one at zero is left, as shown in Fig.~\ref{fig1}(b).  When $\Omega>4E_r$, two minimum points in the middle band merge into one, as shown in Fig.~\ref{fig1}(c).  When the scattering length $a$ increases, the bound states have less binding energy and all the bands are lifted up,  as shown in Fig.~\ref{fig1}(d).

\begin{figure}[]
	\includegraphics[width=\columnwidth]{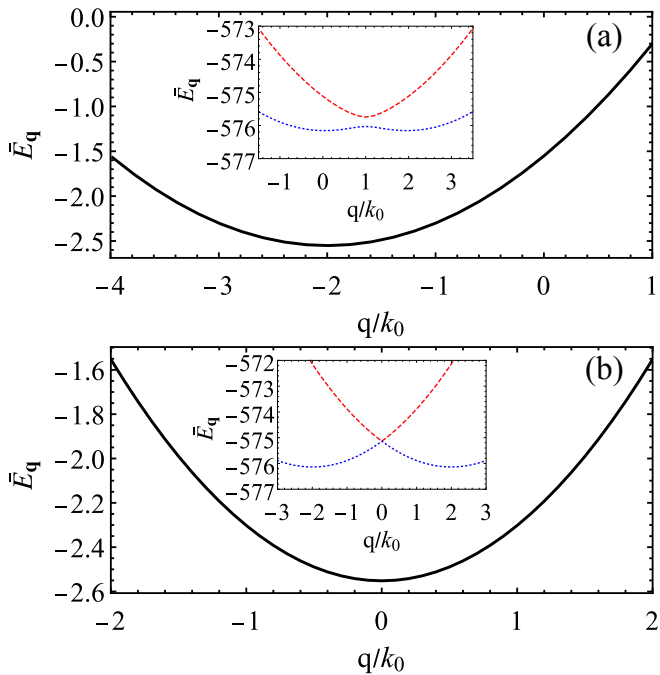}
	\caption{\label{fig2}Bound-state bands with asymmetric interactions at $\Omega=0.4E_r$ and $1/(k_0a_{\downarrow\!\downarrow})=24$. (a) For $a_{\downarrow\!\downarrow}=a_{\uparrow\!\downarrow}=100a_0$, and $ a_{\uparrow\!\uparrow}=1500a_0$, with the Bohr radius $ a_0 $, the top band is shown with the minimum located near $-2\mathbf{k}_0$ and the bottom two bands is shown in the inset. (b) For $a_{\downarrow\!\downarrow}=a_{\uparrow\!\uparrow}=100a_0$, and $ a_{\uparrow\!\downarrow}=1500a_0$, the top band minimum is now located at zero momentum, while the bottom band minimum are located near $\pm 2\mathbf{k}_0$ as shown in the inset.}
\end{figure}

The bound-state bands also change with the asymmetry of interactions.  In an extreme case with $a_{\downarrow\downarrow}=a_{\uparrow\downarrow} \ll a_{\uparrow\uparrow}$, one band is lifted up with the band minimum located near COMM $-2\hbar\mathbf{k}_0$, as shown in Fig.~\ref{fig2}(a).  In another case with with $a_{\uparrow\uparrow}=a_{\downarrow\downarrow} \ll a_{\uparrow\downarrow}$, one band is raised up with the band minimum located at zero COMM , as shown in Fig.~\ref{fig2}(b).  In these two cases, the middle band has only one minimum while the bottom band has one or two minima depending on the Raman strength $\Omega$. The minimum energies of these two lower bands are almost the same as the case of symmetric interactions with the same $a_{\uparrow\downarrow}$.

The bound-state wavefunctions can also be solved~\footnote[1]{One can obtain the wavefunction from $ \Phi_\mathbf{kq}=M_\mathbf{kq}^{-1}\Gamma_i $, where $ \Gamma_i $ is the $ i $-th eigenvector of $ (\mathbf{G}\frac{1}{V}\sum_{\mathbf{k}}M^{-1}_\mathbf{kq}-\mathbf{I}) $ with eigenvalue $ \lambda_i(E_{2\mathbf{q}})=0 $. }.  In the case with asymmetric interactions $a_{\uparrow\!\uparrow} \gg a_{\downarrow\!\downarrow}=a_{\uparrow\!\downarrow}$, we find that at the bottom of top band, the bound state is largely made of atom pairs with spin-$\uparrow\uparrow$ when the Raman strength is weak, as shown in Table~\ref{table}.  When the Raman strength increases to the resonance point where the bound state energy equals to twice the lowest atom energy, the bound state consists of atom pairs with all the spin configurations.

\begin{table}[]
	\caption{\label{table}Bound states of the first band in the case of $ a_{\uparrow\!\uparrow}=1500a_0, a_{\downarrow\!\downarrow}=a_{\uparrow\!\downarrow}=100a_0 $ and $1/(k_0a_{\downarrow\!\downarrow})=24$. The first two columns are in unit of $ E_r $. The momentum is in unit of $ \hbar\mathbf{k}_0 $. }
	\begin{ruledtabular}
		\begin{tabular}{ccccccc}
			$\Omega$  & $ E_{2\mathbf{q}}-2E_\text{min} $   & $ 2\mathbf{q} $  &  $ \sum_{\mathbf{k}}|\phi_{\uparrow\!\uparrow}|^2 $  &  $ \sum_{\mathbf{k}}|\phi_{\downarrow\!\downarrow}|^2 $ & $ \sum_{\mathbf{k}}|\phi_{\uparrow\!\downarrow}|^2 $  \\ \hline
			2 & -4.682 & -2 & 0.995 & 0.00003 & 0.00234 \\
			2 & -2.695 & 0 & 0.987 & 0.0004 & 0.0063 \\
			6.9 & -1.08	& -2 & 0.924 & 0.00548 & 0.0354 \\
			6.9 & -0.811	& -1 & 0.855 & 0.0186 & 0.0633 	\\
			6.9 & -0.0123 & 0 & 0.373 & 0.196 & 0.216 \\
			7.09 & 0 & 0 & 0.25 & 0.25 & 0.25 \\
		\end{tabular}
	\end{ruledtabular}
\end{table}

\textit{Induced resonance}.   At zero Raman strength, the resonance condition is the same as that without SOC, i.e. when the scattering length diverges.  At finite Raman strength, the resonance condition changes due to the reconstruction of bound-state bands.  In experiments, atoms are often condensed in the two single-particle states with the lowest energy.  Since the different spin channels are coupled at finite Raman strength, the resonance occurs whenever the bound state energy satisfies $E_{\pm2\mathbf{q}_0}=2E_{\min}$ or $E_{0}=2E_{\min}$. As a result, multiple resonances can be induced at finite scattering lengths with finite Raman strength.

\begin{figure}[]
	\includegraphics[width=\columnwidth]{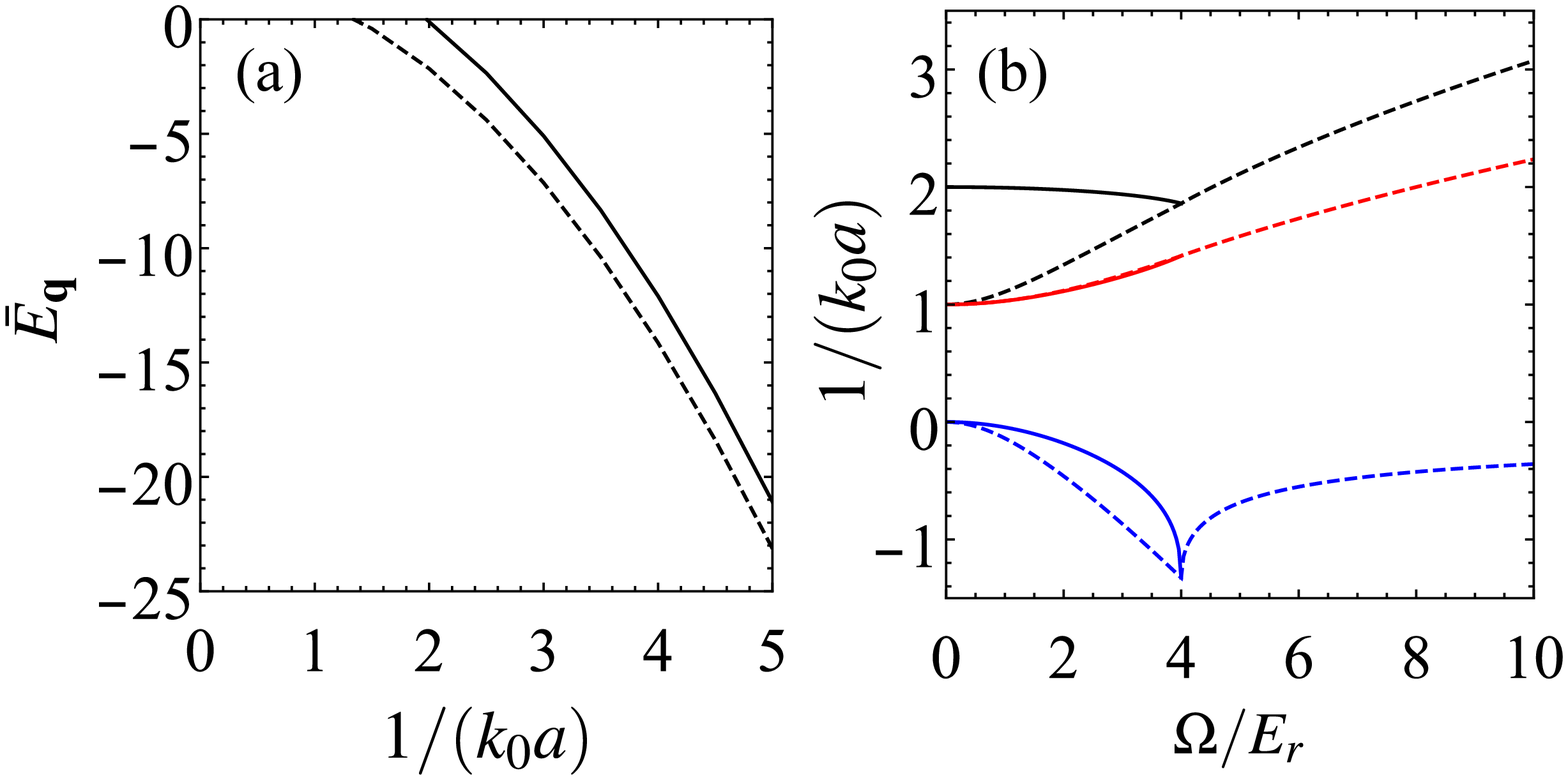}
	\caption{\label{fig3} Induced resonance with symmetric interactions.  (a) Bound-state energy in the top band as a function of scattering length $a$, where $ \Omega=2E_r $.  The solid and dashed lines are dimensionless bound state energy $\bar{E}_\mathbf{q}= (E_\mathbf{q}-2E_{\min})/(2E_r) $ at $\mathbf{q}=2\mathbf{q}_0$ and $0$ respectively.  (b) Scattering length $a$ vs Raman strength $\Omega$ at resonances.  The solid and dashed lines are for COMM  $2\hbar\mathbf{q}_0$ and $0$ respectively. The resonances in the middle band (red lines) occur almost simultaneously for $ \Omega<4E_r $, and the red dashed line is simply described by $ \sqrt{-E_{\min}/E_r} $.}
\end{figure}

For symmetric interactions $ a_{\uparrow\!\uparrow}= a_{\downarrow\!\downarrow}=a_{\uparrow\!\downarrow}=a$, as shown in Fig.~\ref{fig3}(a) and (b) , resonance condition $E_{2\mathbf{q}_0}=2E_{\min}$ and $E_0=2E_{\min}$ are satisfied in the top band at two different scattering lengths for $\Omega<4E_r$.  There are totally three different resonances that one can induce by tuning Raman strength $\Omega$ for each band, but due to symmetry $E_{2\mathbf{q}_0}=E_{-2\mathbf{q}_0}$, they are located at only two different scattering lengths.  When $\Omega>4E_r$, there is only one induced resonance for each band as $\mathbf{q}_0=0$.

\begin{figure}[]
	\includegraphics[width=\columnwidth]{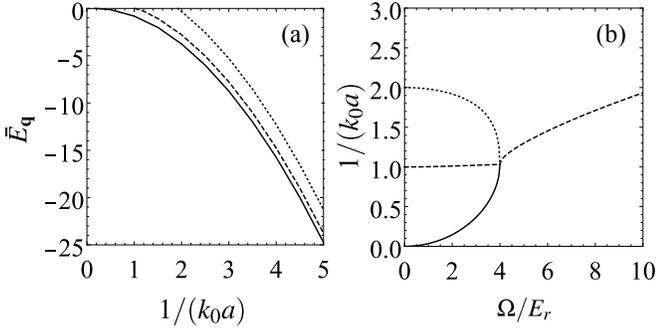}
	\caption{\label{fig4} Induced resonance with asymmetric interactions $a_{\uparrow\!\uparrow} \gg a_{\downarrow\!\downarrow}=a_{\uparrow\!\downarrow}$ and $1/(k_0a_{\uparrow\!\downarrow})=24 $.  (a) Bound-state energy in the top band as a function of scattering length $ a_{\uparrow\!\uparrow}$, where $ \Omega=2E_r $.  The solid, dashed and dotted lines are dimensionless bound state energy $\bar{E}_\mathbf{q}= (E_\mathbf{q}-2E_{\min})/(2E_r) $ at $\mathbf{q}=-2\mathbf{q}_0,0,2\mathbf{q}_0$ respectively.    (b) Scattering length $a$ vs Raman strength $\Omega$ at resonances in the top band.  The solid, dashed and dotted lines are for COMM  $-2\hbar\mathbf{q}_0$, $0$, and $2\hbar\mathbf{q}_0$ respectively.}
\end{figure}

With asymmetric interactions $a_{\uparrow\!\uparrow} \gg a_{\downarrow\!\downarrow}=a_{\uparrow\!\downarrow}$,
for fixed $ a_{\downarrow\!\downarrow} $ and $ a_{\uparrow\!\downarrow} $, the bound-state energy displays different behavior as a functions of $ 1/(k_0a_{\uparrow\!\uparrow}) $ in different bands. When $a_{\uparrow\!\uparrow}$ increases, only the top bound-state band can reach the lowest scattering energy $2E_{\min}$.  The other two bands are insensitive to the change in $a_{\uparrow\!\uparrow}$ as shown in Fig.~\ref{fig2}(a).  When $\Omega<4E_r$, as shown in Fig.~\ref{fig4}, there are three induced resonances with COMM  $0,\pm2\hbar\mathbf{q}_0$. When $\Omega>4E_r$, there is only one induced resonance.

\textit{Effective interactions near Resonance}.  With finite Raman strength, the effective interactions between atoms are no longer described by the bare coupling constants, but given by the $T$-matrix which satisfying Bethe-Salpeter equation
\begin{equation}\label{Teq}
\mathbf{T}=\mathbf{G'}+\mathbf{G'}\mathbf{\chi}\mathbf{T},
\end{equation}
where $\mathbf{G'}$ is the 3 by 3 coupling matrix and $\mathbf{\chi}$ is the pair susceptibility function.
The solution of Eq.~\eqref{Teq} is given by
\begin{equation}\label{T-matrix}
\mathbf{T}^{-1}=-\dfrac{m}{4\pi\hbar^2}
\begin{pmatrix}
A_1-\dfrac{1}{a_{\uparrow\!\uparrow}} & C_3 & \sqrt{2}C_1 \\
C_3 & A_2-\dfrac{1}{a_{\downarrow\!\downarrow}} & \sqrt{2}C_2 \\
\sqrt{2}C_1  & \sqrt{2}C_2 & A_3-\dfrac{1}{a_{\uparrow\!\downarrow}}
\end{pmatrix}.
\end{equation}

With symmetric interactions $a_{\uparrow\!\uparrow}=a_{\downarrow\!\downarrow}=a_{\uparrow\!\downarrow}=a$, the coupling matrix $\mathbf{G'}$ is proportional to the identity matrix and Eq.~\eqref{T-matrix} can be rewritten as,
\begin{equation}\label{Teq1}
\mathbf{T}^{-1}=\frac{m}{4\pi\hbar^2}(\frac{1}{a}\mathbf{I}-\mathbf{\chi}'),
\end{equation}
where $\mathbf{I}$ is the identity matrix and $\mathbf{\chi}'$ is the modified pair susceptibility function.
From Eq.~\eqref{Teq1}, the $T$-matrix can be solved explicitly,
\begin{equation}
\mathbf{T}^{-1}=\frac{m}{4\pi\hbar^2}\mathbf{U}(\frac{1}{a}\mathbf{I}-\mathbf{\lambda})\mathbf{U}^{-1},
\end{equation}
where $\mathbf{\lambda}$ and $\mathbf{U}$ are eigenvalue and unitary transformation matrices of $\mathbf{\chi}'$, $\mathbf{\chi}'=\mathbf{U} \mathbf{\lambda} \mathbf{U}^{-1}$, $\lambda_{ij}=\delta_{ij}(1/a^{i}_\text{res})$ and $1/a^{i}_\text{res}$ is the $i$-th eigenvalue of $\mathbf{\chi}'$.
A resonance occurs whenever the scattering length $a$ is near $a^{i}_\text{res}$.  The effective interactions near a resonance $a=a^{i}_\text{res}$ can be approximated by
\begin{equation}
\mathbf{T}_{l}^{m}\approx\frac{4\pi\hbar^2a^{i}_\text{res}a}{m(a^{i}_\text{res}-a)}\mathbf{U}_{l}^{i}\mathbf{U}_{m}^{i},
\end{equation}
where $ l,m,i $ refer to the scattering channels $\uparrow\!\uparrow$, $ \downarrow\!\downarrow $ and $ \uparrow\!\downarrow $.

With asymmetric interactions $a_{\uparrow\!\uparrow} \gg a_{\downarrow\!\downarrow}\approx a_{\uparrow\!\downarrow}$, we obtain each matrix element of $T$ in the leading order of $a_{\uparrow\!\downarrow}$ as given by
\begin{align}
\nonumber
T_{\uparrow\!\uparrow}^{\uparrow\!\uparrow}&=\frac{4\pi\hbar^2}{m}(\frac{1}{a_{\uparrow\!\uparrow}}-{\chi'}_{\uparrow\!\uparrow}^{\uparrow\!\uparrow})^{-1} ,\\
\nonumber
T_{l}^{l}&=\frac{4\pi\hbar^2}{m}a_{l} ,\\
\nonumber
T_{l}^{\uparrow\!\uparrow}&=T_{\uparrow\!\uparrow}^{\uparrow\!\uparrow}{\chi'}_{\uparrow\!\uparrow}^{l}a_{l} ,\\
T_{\downarrow\!\downarrow}^{\uparrow\!\downarrow}&=T_{\uparrow\!\uparrow}^{\uparrow\!\uparrow}\left[{\chi'}_{\uparrow\!\downarrow}^{\downarrow\!\downarrow}(\frac{1}{a_{\uparrow\!\uparrow}}-{\chi'}_{\uparrow\!\uparrow}^{\uparrow\!\uparrow})+{\chi'}_{\uparrow\!\uparrow}^{\downarrow\!\downarrow}{\chi'}_{\uparrow\!\downarrow}^{\uparrow\!\uparrow}\right]a_{\downarrow\!\downarrow}a_{\uparrow\!\downarrow} ,
\end{align}
where $ l $ refers to either the scattering channel $ \downarrow\!\downarrow $ or $ \uparrow\!\downarrow $. To the leading order of $a_{\uparrow\!\downarrow}$, the dominant interaction near resonance is in the ${\uparrow\!\uparrow}$ channel, $ \mathbf{T}_{\uparrow\!\uparrow}^{\uparrow\!\uparrow}=\frac{4\pi\hbar^2}{m}(\frac{1}{a_{\uparrow\!\uparrow}}-A_1)^{-1} $ and the resonance occurs at $a_{\uparrow\!\uparrow}=A_1^{-1}$ where $A_1={\chi'}_{\uparrow\!\uparrow}^{\uparrow\!\uparrow}$ as given in Eq. \eqref{AC}.

\textit{Discussion and Conclusion}.
In experiments, ${}^{87}$Rb atom gas with Raman-induced SOC is a common platform for studying spin-orbit coupled bosons~\cite{PhysRevA.87.042514}. Usually, a pair of Raman laser beams couple two hyperfine states, $ \ket{\uparrow}=\ket{F=1,m_F=0} $ and $ \ket{\downarrow}=\ket{F=1,m_F=-1} $. The scattering lengths of different channels are almost equal $ a_{\uparrow\!\uparrow}\approx100.8a_0, a_{\downarrow\!\downarrow}=a_{\uparrow\!\downarrow}\approx100.4a_0 $, with $1/(k_0a_{\uparrow\!\downarrow}) \approx 24$ for Raman laser wavelength 804.1~nm~\cite{PhysRevLett.88.093201,Nature2011Spielman}.  The Raman strength can be tuned up to the order of $10E_r$.  At $\Omega=10E_r$, we find for symmetric interactions the resonance positions is at $a=780a_0$, much larger than the scattering lengths.  It is difficult to observe the induced resonance by adjusting SOC alone.  However the scattering length $ a_{\uparrow\!\uparrow} $ is tunable by Feshbach resonance~\cite{PhysRevLett.89.283202}.  It is possible to observe the induced resonance in the region $ a_{\uparrow\!\uparrow} \gg  a_{\uparrow\!\downarrow}\approx a_{\downarrow\!\downarrow} $. For example, at $\Omega=10E_r$, the resonance position is given by $ a_{\uparrow\!\uparrow}=1242a_0 $, for $ 1/(k_0a_{\uparrow\!\downarrow})=24 $ shown in Fig.~\ref{fig4}(b), available by Feshbach resonance in experiments.  When $\Omega<4E_r$, three resonances can be observed instead of one.

The resonance induced by SOC has important effects in a Bose gas.  Near the resonance, severe particle loss is expected to occur which may be used as a tool to locate the resonance.  Over the resonance, the effective interaction turns into an attractive interaction, and the system is expected to collapse at low temperatures.  The effective interaction is strongly modified by SOC near the resonance, and interaction-determined many-body properties are expected to be different from predictions of the simple mean-field theory, which will be the subject of our future work.

In summary, we study bound state bands and resonances in a Bose gas with SOC.  We find that finite Raman strength generates coupling among different scattering channels, leading to the reconstruction of bound-state bands.  The resonance positions are also shifted due to finite Raman coupling strength, and the effective interactions near these resonance are obtained.  We predict that by tuning the scattering length in one intra-species channel, the resonance induced by SOC can be observed in rubidium-87 systems.

\begin{acknowledgements}
We would like to thank X.-J. Zhou, Z.-Q. Yu, P. Zhang, and T.-L. Ho for helpful discussions.  This work is supported by the National Key Research and Development Project of China under Grant No. 2016YFA0301501.
\end{acknowledgements}

\bibliography{manuscripts}
\end{document}